\documentclass{article}
\usepackage{spconf,amsmath,amssymb,graphicx,booktabs,enumitem}
\usepackage[hidelinks]{hyperref}
\usepackage{algorithm,algpseudocode,cite}
\usepackage{threeparttable}
\usepackage{comment}
\usepackage[table]{xcolor}

\newcommand{\para}[1]{\par\addvspace{2pt}\noindent\textbf{#1:}}
\def\xhat{\hat{\mathbf{x}}}
\def\xmod{\tilde{\mathbf{x}}}

\title{IMPROVING PREDICTED MOS SCORES, NOT PERCEIVED QUALITY:\\
MULTI-PREDICTOR TEST-TIME OPTIMIZATION OF ENHANCED SPEECH}

\name{\shortstack[c]{Tsubasa Ochiai$^1$, Marc Delcroix$^1$, Nahomi Kusunoki$^2$, Rintaro Ikeshita$^1$,\\
Naohiro Tawara$^1$, Naoyuki Kamo$^1$, Tetsuji Ogawa$^2$, Shoko Araki$^1$}}
\address{
  $^1$NTT, Inc., Japan \ \ \ $^2$Waseda University, Japan
}

\begin{document}
\ninept
\makeatletter
\def\setdisp{%
  \setlength\abovedisplayskip{6pt plus 2pt minus 2pt}%
  \setlength\belowdisplayskip{6pt plus 2pt minus 2pt}%
  \setlength\abovedisplayshortskip{0pt plus 2pt}%
  \setlength\belowdisplayshortskip{4pt plus 2pt minus 2pt}}
\g@addto@macro\normalsize{\setdisp}
\setdisp
\makeatother
\maketitle

\begin{abstract}
Non-intrusive MOS predictors are widely used instead of subjective listening tests to evaluate and rank speech enhancement (SE) systems. If they accurately reflect perceived quality, raising their scores should lead to higher-quality speech.
We present the first comprehensive analysis of test-time optimization for the SE task, which directly modifies the enhanced signal to raise the average of multiple MOS predictor scores.
On seven systems from the URGENT 2026 challenge, we find that 1)~all the optimized predicted scores increase while reference-based metrics remain nearly unchanged, 2)~a non-optimized predicted score does not increase, and 3)~a MUSHRA listening test shows no improvement in perceived quality. 
These findings reveal a risk that such optimization can distort evaluations, e.g., biasing comparisons of SE systems regardless of their perceived quality.
We believe these findings can inform future evaluation practices: they suggest that predictors used for optimization should not be used for evaluation, and that challenges should keep the predictors used for ranking undisclosed.
\end{abstract}

\begin{keywords}
speech enhancement, speech quality assessment, test-time optimization, challenge evaluation
\end{keywords}

\section{Introduction}
\label{sec:intro}

Subjective listening tests are generally regarded as the most reliable method
for judging the quality of speech enhancement (SE) systems, but they are
time-consuming and financially costly. Therefore, the scores of non-intrusive mean opinion score (MOS) predictors~\cite{dnsmos,nisqa,utmos,scoreq} are widely used as an alternative. These predictors take a single speech signal as input, without any reference to the corresponding clean speech, and predict
the MOS that listeners would assign to it. These predicted scores are used to compare systems in research papers and contribute to
the official rankings of challenges~\cite{urgent2026,realtse2026,avse2026}.

If these predictors accurately reflect the speech quality perceived by human listeners,
raising their scores should produce better speech. 
With this expectation, some studies train SE systems with an additional loss term to increase the predicted scores (e.g., \cite{multimetric2025,matsunaga2026three}).
We could hold the same expectation at test time.
Since the predictors need no reference signal, the enhanced signal can be directly modified for each test utterance to increase its predicted scores, without retraining the SE system. 
We refer to this procedure as test-time optimization.
In this paper, we address the following research question: Does increasing predicted MOS scores by the test-time optimization improve the quality that listeners actually perceive?

The answer matters in both directions. 
If increases in the predicted scores correspond to increases in the quality perceived by listeners, test-time optimization would be an effective post-processing step that improves the enhanced signals of any SE system.
If they do not, the same procedure may become an adversarial attack~\cite{fgsm,lin2022} on evaluations that rely on these predictors, because the reported scores and the rankings computed from them could be raised without any improvement in perceived quality. 
A predicted MOS score is often used as one of the criteria to assess improvement, and a score that can be raised on demand provides less convincing evidence of actual improvement.
Which of the two cases holds remains unknown: to the best of our knowledge, no study has examined whether the test-time optimization of the enhanced speech to increase the predicted MOS scores improves the perceived quality.

To answer this question, we conduct a systematic evaluation of test-time optimization with multiple
MOS predictors, applied to the enhanced signals of the URGENT 2026 challenge.
We jointly optimize the enhanced signal for several predictors.
Achieving high scores on several predictors, which differ in architecture and training data, can be a stronger constraint than on one, and a signal achieving such scores can be expected to have higher perceived quality.
Conversely, if the scores of several predictors can be raised simultaneously without any improvement in
perceived quality, this would demonstrate a risk for evaluations that use such predictors, including challenges.
We analyze the effect of this test-time optimization from multiple perspectives: 1)~the optimized MOS predictors in the objective, 2)~an unseen MOS predictor outside the objective, 3)~various reference-based metrics, 4)~the challenge ranking, and 5)~the ratings of human listeners in a MUSHRA test. 
We believe that such an analysis is important for guiding evaluation practices in future SE research. 
Our main contributions and new findings are as follows:

\begin{itemize}
[leftmargin=1.2em,itemsep=2pt,topsep=2pt,parsep=0pt]
\item We introduce test-time optimization that raises the average of multiple MOS predictor scores, and \emph{conduct a comprehensive analysis on seven systems from the URGENT 2026 challenge}, measuring its effect on the challenge ranking and human listeners.
\item Test-time optimization simultaneously increases the scores of all the MOS predictors in the objective, while the reference-based metrics remain nearly unchanged.
However, \emph{these increases do not transfer to an unseen MOS predictor}.
\item In a MUSHRA listening test, raising the predicted scores by test-time optimization does not lead to better perceived quality: \emph{despite the large increase in the predicted scores, listeners rate the optimized signals no better than the original ones}.
\item \emph{Test-time optimization can also distort the outcome of a challenge}, moving systems several places in the ranking and potentially displacing other participants. We therefore recommend not evaluating SE systems with the predictors used for optimization, and keeping the predictors used for ranking undisclosed.
\end{itemize}

\section{Related work}
\label{sec:related}

Lin \textit{et al.}~\cite{lin2022} showed that DNSMOS can be driven to a chosen target score by an adversarial perturbation constrained to lie more than $30$~dB below the signal level, and confirmed in an AB test that listeners could not distinguish the perturbed signal from the original.
Our objective in Section~\ref{sec:method} also combines a term that raises the predicted score with a term that penalizes the deviation from the original signal, but it places no explicit bound on the amplitude of the added signal, allowing the optimization to potentially improve audio quality.
Concurrently and independently, Huang and Toda~\cite{huang2026} report similar attacks on UTMOS. Both studies perturb clean or noisy speech rather than the output of an SE system.
In contrast to these studies, we conduct a comprehensive analysis of test-time optimization for the SE task, measuring the effect of jointly optimizing for multiple predictors on 1)~an unseen predictor, 2)~various reference-based metrics, 3)~the challenge ranking, and 4)~the ratings of human listeners. 

In another concurrent and independent study, Klement \textit{et al.}~\cite{merl2026} applied a technique similar to that of Lin \textit{et al.}~\cite{lin2022} to the recent REAL-TSE challenge~\cite{realtse2026}.
They perturbed their own submission and raised DNSMOS from $2.83$ to $4.18$, while the task-specific metrics remained unchanged.
However, they did not evaluate the perturbed signals in a listening test, leaving open whether perceived quality improved.
The recent use of such an attack in a real challenge underlines the need for a dedicated analysis, which this work provides.

%\vspace{-1mm}
\section{Test-time optimization of enhanced speech}
\label{sec:method}

\vspace{-1mm}
\subsection{Preliminaries}

Let $\mathbf{y} \in \mathbb{R}^{T}$ denote a $T$-length time-domain waveform of the observed signal of one test utterance.
Given $\mathbf{y}$ as input, the enhanced signal $\xhat \in \mathbb{R}^{T}$ and its predicted MOS score by the $m$-th MOS predictor $\hat{m}_{m} \in \mathbb{R} \ (m = 1, \dots, M)$ are obtained as
\begin{equation}
  \xhat = \mathrm{SE}(\mathbf{y}; \Theta), \qquad
  \hat{m}_{m} = \mathrm{pMOS}_{m}(\xhat; \Phi_{m}),
  \label{eq:pipeline}
\end{equation}
where $\mathrm{SE}(\cdot\,; \Theta)$ denotes the functional representation of an SE system with parameters $\Theta$, and $\mathrm{pMOS}_{m}(\cdot\,; \Phi_{m})$ denotes that of the $m$-th MOS predictor with parameters $\Phi_{m}$ used for evaluation.

\vspace{-1mm}
\subsection{Optimization objective}

Test-time optimization directly optimizes the enhanced signal while keeping the SE system fixed.
By adding a signal $\mathbf{z} \in \mathbb{R}^{T}$, we modify the enhanced signal $\xhat$ as
\begin{equation}
  \xmod = \xhat + \mathbf{z},
  \label{eq:add}
\end{equation}
where $\xmod \in \mathbb{R}^{T}$ denotes the modified signal. 
We optimize $\mathbf{z}$ to increase the average of the $M$ predicted scores while penalizing the deviation of $\xmod$ from $\xhat$,
\begin{equation}
  \hat{\mathbf{z}} = \operatorname*{arg\,min}_{\mathbf{z}}\; (1-\lambda)\, D(\xmod, \xhat) - \frac{\lambda}{M} \sum_{m=1}^{M} \mathrm{pMOS}_m(\xmod; \Phi_{m}),
  \label{eq:obj}
\end{equation}
where $D(\cdot, \cdot)$ denotes a distance between two signals and $\lambda \in [0, 1]$
denotes a weight that controls the trade-off between the two terms. We optimize
only the added signal $\mathbf{z}$, whereas $\Theta$ and $\Phi_{m}$ remain
fixed. 
The enhanced signal is therefore modified to increase the predicted scores.
If the predictors reflect perceived quality, the modified signal
should sound better than the original. Note that the optimization is
carried out independently for each utterance, without using a reference signal
or access to the SE system that produced $\xhat$.

\vspace{-1mm}
\subsection{Optimization procedure}

Algorithm~\ref{alg:tto} summarizes the procedure of the test-time optimization of enhanced speech.
We initialize $\mathbf{z}$ with a signal of small amplitude and update it by gradient descent to minimize the objective in Eq.~\eqref{eq:obj} for a fixed number of iterations $N$ with learning rate $\eta_{i}$.
At each iteration, the gradients are backpropagated through the predictors to $\mathbf{z}$, and only $\mathbf{z}$ is updated while the predictors remain unchanged.

\begin{algorithm}[t]
\caption{Test-time optimization of enhanced speech}
\label{alg:tto}
\small
\begin{algorithmic}[1]
\Require enhanced signal $\xhat$, predictors $\mathrm{pMOS}_{m}$, weight $\lambda$
\State define $\mathcal{L}(\mathbf{z}) = (1-\lambda)\, D(\xhat + \mathbf{z}, \xhat)
        - \frac{\lambda}{M} \sum_{m=1}^{M} \mathrm{pMOS}_{m}(\xhat + \mathbf{z})$
\State initialize $\mathbf{z}$ with small random values
\For{$i = 1$ to $N$}
  \State $\mathbf{z} \gets \mathbf{z} - \eta_{i} \nabla_{\mathbf{z}} \mathcal{L}(\mathbf{z})$
\EndFor
\State \Return $\xmod = \xhat + \mathbf{z}$
\end{algorithmic}
\end{algorithm}

\vspace{-1mm}
\section{Experimental design}
\label{sec:setup}

In this paper, we evaluate the effect of test-time optimization on a recent SE
challenge, i.e., URGENT 2026~\cite{urgent2026}, using the systems submitted to the
actual challenge. 
Through this evaluation, we investigate how jointly optimizing for multiple MOS predictors affects the non-intrusive and the reference-based metrics, the quality that listeners perceive, and the official ranking of the challenge.

\vspace{-1mm}
\subsection{Data and evaluation}

\para{Data and systems} Our analysis is based on the non-blind test set, not on the blind test set used for the final ranking of the challenge, including its subjective listening test.
The non-blind test set of URGENT 2026 Track~1 contains $1023$ utterances, covering various sampling rates, languages, and distortion types. 
We use this set because it is publicly available with reference audio and labels, whereas the blind test set is not. We analyze seven systems: the six that reached the subjective stage of the challenge and the BSRNN baseline. 
We refer to the six by their leaderboard names, i.e., WR\cite{rong2026gap}, subatomicseer\cite{goswami2026hybrid}, baird\cite{matsunaga2026three}, GHW, Ali-Universal-SE\cite{liu2026hybrid}, and RICK2000, citing the system description papers published in the ICASSP 2026 special session on URGENT.
The enhanced signals of these systems are not public and were
provided to us by the organizers under the challenge rules. The BSRNN baseline
model is publicly available, and therefore, our results for the baseline on the
non-blind test set can be reproduced from public data alone.

\para{Evaluation metrics} We report all $12$ metrics of the
URGENT 2026 non-blind leaderboard, computed with the evaluation script of the
challenge~\cite{urgentcode}, and reimplement its ranking
procedure~\cite{urgent2026}. 
The metrics are grouped into four categories: the non-intrusive metrics
(DNSMOS~\cite{dnsmos}, NISQA~\cite{nisqa}, UTMOS~\cite{utmos}, and
SCOREQ~\cite{scoreq}), the intrusive metrics (PESQ~\cite{pesq} and ESTOI~\cite{estoi}), the
downstream-task-independent metrics (SpeechBERTScore~\cite{sbertscore} and LPS~\cite{lps}), and the
downstream-task-dependent metrics (SpkSim, EmoSim, LID, and CAcc [\%]). 
All metrics other than the non-intrusive ones are reference-based, which are computed
against the clean reference or the ground-truth labels (see~\cite{urgent2026}
for details). We additionally evaluate Distill-MOS~\cite{distillmos} as an unseen predictor outside the objective. It is not part of the challenge and does not enter the ranking,
although we report it with the other metrics.

\para{Ranking} Each metric is rounded to two decimals and dense-ranked
across systems, where systems with the same rounded value share the same rank and the
subsequent rank is not skipped. The ranks are then averaged within
each of the four categories, and the mean of the four category scores is dense-ranked to give the final ranking. 
When applied to the scores of all 24 entries on the leaderboard, our implementation reproduces their overall scores and the ranks.
For each SE system, we replace its entry with the metrics recomputed on its modified signals, leaving the other 23 entries unchanged.
Any change in the ranking is therefore caused only by the optimization.

\para{Subjective listening test} We conducted a MUSHRA-style listening test
(i.e., ITU-R Recommendation BS.1534-3~\cite{bs1534}) on $10$ utterances drawn at random from the English $48$~kHz subset of the
non-blind test set. 
We adopted a MUSHRA test (0--100) rather than a conventional MOS test (1--5), because it presents all conditions of an utterance side by side, allowing each enhanced signal to be compared directly with its modified version.
For each utterance, the clean signal serves as the
reference and as the hidden reference, and the noisy mixture serves as the
anchor. 
Twelve listeners working on speech processing participated in the
test, using the same laptop and headphones in the same room. 
One of them rated the hidden reference below $90$ on half of the utterances, and was therefore excluded by the post-screening criterion, which leaves $11$ listeners.
For each utterance, each listener rated five signals: the original submissions from ``baird'' and ``RICK2000'', the modified signals of both systems optimized with $\lambda = 0.005$, and the modified signal of ``baird'' optimized with $\lambda = 0.05$.
We included the larger $\lambda$ for one system only, to reduce the burden on the listeners.
We chose these two systems because they ranked first and last in the final subjective listening test of the challenge.

\vspace{-1mm}
\subsection{Optimization}

We optimize the four non-intrusive MOS predictors of the challenge, i.e.,
DNSMOS, NISQA, UTMOS, and SCOREQ. DNSMOS returns SIG, BAK, and OVRL, and we
optimize the mean of the three, while the ranking uses OVRL alone. To obtain
gradients, we replaced the non-PyTorch operations of each predictor with
PyTorch ones. Because these replacements are not always exact (e.g.,
resampling), the added signal $\mathbf{z}$ is optimized against a slightly different
function from the one that scores it. 
The added signal has the same sampling rate as $\xhat$, and each predictor resamples $\xmod$ to its own operating rate inside the differentiable graph, following the evaluation code.

We adopt the multi-resolution L1 loss of the ESPnet implementation~\cite{multiresl1} as the distance
$D(\cdot, \cdot)$ in Eq.~\eqref{eq:obj}. 
It combines a time-domain L1 loss with those on STFT magnitudes at window sizes $256$, $512$, $768$, and $1024$.
We initialize $\mathbf{z}$ with Gaussian noise of standard
deviation $1\times10^{-4}$. We then run $N = 500$ iterations of gradient descent, starting
from a learning rate of $5\times10^{-5}$ and decaying it by a factor of $0.99$
at each iteration.
%weighted $0.5$

\vspace{-1mm}
\section{Results and discussion}
\label{sec:results}

\subsection{Objective evaluation}
\label{ssec:obj}

\begin{table*}[t]
\centering
\setlength{\belowcaptionskip}{7pt}
\caption{Scores and ranks of the enhanced signal $\xhat$ (shaded rows), and score differences and rank changes between the modified signal $\xmod$ and $\xhat$. The enhanced signal is jointly optimized for the four predictors, except in the rows marked (S), where it is optimized for DNSMOS alone.}
%\caption{Scores and ranks of the enhanced signal $\xhat$ (shaded rows), and score differences and rank changes between the modified signal $\xmod$ and the enhanced signal $\xhat$. The enhanced signal is jointly optimized for the four predictors, except in the rows marked (S), where it is optimized for DNSMOS alone.}
%The four predictors are jointly optimized, except in the rows marked (S), where DNSMOS is optimized alone.}
\vspace{5pt}
\label{tab:main}
\setlength{\tabcolsep}{1.98pt}
\setlength{\aboverulesep}{0pt}
\setlength{\belowrulesep}{0pt}
\renewcommand{\arraystretch}{1.15}
\footnotesize
\begin{tabular}{ll|rrrrr|rr|rr|rrrr|cc}
\toprule
\rule{0pt}{2.62ex} & & \multicolumn{4}{c}{Non-intrusive (Optimized)} & Unseen & \multicolumn{2}{c|}{Intrusive} & \multicolumn{2}{c|}{Task-indep.} & \multicolumn{4}{c|}{Task-dependent} & \multicolumn{2}{c}{Rank} \\
%\rule{0pt}{2.62ex} & & \multicolumn{4}{|c}{Non-intrusive (Optimized)} & Unseen & \multicolumn{2}{|c}{Intrusive} & \multicolumn{2}{|c}{Task-indep.} & \multicolumn{4}{|c}{Task-dependent} & \multicolumn{2}{|c}{Rank} \\
System\rule[-1.08ex]{0pt}{0pt} & $\lambda$ & DNSMOS & NISQA & UTMOS & SCOREQ & Distill & PESQ & ESTOI & SBERT & LPS & SpkSim & EmoSim & LID & CAcc & non-int. & overall \\
\midrule
\rowcolor{gray!15}
WR & $-$ & $3.06$ & $3.37$ & $2.77$ & $3.67$ & $3.68$ & $2.81$ & $0.86$ & $0.89$ & $0.83$ & $0.76$ & $0.99$ & $0.96$ & $88.16$ & 3 & 1 \\
 & $0.005$ & $+0.24$ & $+1.19$ & $+0.15$ & $+0.12$ & $+0.02$ & $0.00$ & $0.00$ & $0.00$ & $0.00$ & $0.00$ & $0.00$ & $0.00$ & $+0.05$ & 3$\to$1 & 1$\to$1 \\
 & $0.05$ & $+0.50$ & $+1.77$ & $+0.54$ & $+0.28$ & $+0.01$ & $-0.06$ & $-0.01$ & $-0.01$ & $0.00$ & $0.00$ & $0.00$ & $0.00$ & $+0.43$ & 3$\to$1 & 1$\to$1 \\
\rowcolor{gray!15}
subatom. & $-$ & $2.96$ & $3.38$ & $2.41$ & $3.32$ & $3.36$ & $2.71$ & $0.84$ & $0.88$ & $0.79$ & $0.76$ & $0.99$ & $0.95$ & $87.76$ & 6 & 2 \\
 & $0.005$ & $+0.49$ & $+1.21$ & $+0.21$ & $+0.19$ & $+0.03$ & $-0.01$ & $0.00$ & $-0.01$ & $0.00$ & $0.00$ & $0.00$ & $0.00$ & $-0.66$ & 6$\to$2 & 2$\to$2 \\
 & $0.05$ & $+0.75$ & $+1.91$ & $+0.69$ & $+0.39$ & $+0.01$ & $-0.17$ & $0.00$ & $-0.02$ & $-0.01$ & $-0.01$ & $0.00$ & $-0.01$ & $-1.34$ & 6$\to$1 & 2$\to$2 \\
\rowcolor{gray!15}
baird & $-$ & $2.96$ & $3.11$ & $2.30$ & $3.24$ & $3.28$ & $2.72$ & $0.85$ & $0.88$ & $0.80$ & $0.77$ & $0.99$ & $0.96$ & $86.90$ & 9 & 3 \\
 & $0.005$ & $+0.49$ & $+1.31$ & $+0.14$ & $+0.14$ & $-0.01$ & $-0.01$ & $0.00$ & $0.00$ & $0.00$ & $0.00$ & $0.00$ & $0.00$ & $-0.22$ & 9$\to$2 & 3$\to$2 \\
 & $0.05$ & $+0.71$ & $+1.99$ & $+0.49$ & $+0.30$ & $-0.02$ & $-0.11$ & $0.00$ & $-0.01$ & $0.00$ & $0.00$ & $0.00$ & $-0.01$ & $-0.31$ & 9$\to$2 & 3$\to$2 \\
\rowcolor{gray!15}
GHW & $-$ & $2.96$ & $3.38$ & $2.64$ & $3.46$ & $3.62$ & $2.69$ & $0.84$ & $0.87$ & $0.77$ & $0.76$ & $0.99$ & $0.94$ & $83.80$ & 4 & 4 \\
 & $0.005$ & $+0.25$ & $+1.04$ & $+0.16$ & $+0.14$ & $+0.01$ & $0.00$ & $0.00$ & $0.00$ & $0.00$ & $0.00$ & $0.00$ & $-0.01$ & $+0.04$ & 4$\to$2 & 4$\to$3 \\
 & $0.05$ & $+0.56$ & $+1.67$ & $+0.56$ & $+0.37$ & $+0.02$ & $-0.05$ & $0.00$ & $-0.01$ & $-0.01$ & $0.00$ & $0.00$ & $-0.01$ & $+0.20$ & 4$\to$1 & 4$\to$4 \\
\rowcolor{gray!15}
Ali-Univ. & $-$ & $3.00$ & $3.25$ & $2.48$ & $3.21$ & $3.35$ & $2.58$ & $0.82$ & $0.88$ & $0.78$ & $0.77$ & $0.99$ & $0.95$ & $86.61$ & 7 & 5 \\
 & $0.005$ & $+0.47$ & $+1.18$ & $+0.16$ & $+0.14$ & $0.00$ & $0.00$ & $0.00$ & $-0.01$ & $0.00$ & $0.00$ & $0.00$ & $0.00$ & $+0.17$ & 7$\to$2 & 5$\to$3 \\
 & $0.05$ & $+0.71$ & $+1.86$ & $+0.56$ & $+0.34$ & $0.00$ & $-0.09$ & $0.00$ & $-0.02$ & $0.00$ & $-0.01$ & $0.00$ & $-0.01$ & $+0.37$ & 7$\to$2 & 5$\to$3 \\
\rowcolor{gray!15}
RICK2000 & $-$ & $2.95$ & $3.28$ & $2.32$ & $3.09$ & $3.32$ & $2.64$ & $0.84$ & $0.87$ & $0.77$ & $0.77$ & $0.99$ & $0.95$ & $85.65$ & 8 & 6 \\
 & $0.005$ & $+0.48$ & $+1.30$ & $+0.20$ & $+0.20$ & $+0.01$ & $-0.01$ & $0.00$ & $0.00$ & $+0.01$ & $0.00$ & $0.00$ & $0.00$ & $+0.16$ & 8$\to$2 & 6$\to$3 \\
 & $0.05$ & $+0.75$ & $+2.00$ & $+0.68$ & $+0.46$ & $-0.02$ & $-0.15$ & $-0.01$ & $-0.02$ & $0.00$ & $-0.01$ & $0.00$ & $-0.01$ & $-0.02$ & 8$\to$2 & 6$\to$4 \\
\rowcolor{gray!15}
Baseline & $-$ & $2.90$ & $2.98$ & $2.21$ & $2.96$ & $3.12$ & $2.47$ & $0.82$ & $0.87$ & $0.77$ & $0.75$ & $0.99$ & $0.94$ & $85.79$ & 14 & 13 \\
 & $0.005$ & $+0.52$ & $+1.31$ & $+0.13$ & $+0.14$ & $0.00$ & $0.00$ & $0.00$ & $0.00$ & $0.00$ & $0.00$ & $0.00$ & $-0.01$ & $-0.31$ & 14$\to$5 & 13$\to$7 \\
 & $0.05$ & $+0.75$ & $+2.06$ & $+0.47$ & $+0.35$ & $-0.01$ & $-0.07$ & $0.00$ & $-0.02$ & $0.00$ & $0.00$ & $0.00$ & $-0.01$ & $-0.09$ & 14$\to$2 & 13$\to$6 \\
\midrule
Baseline (S) & $0.005$ & $+0.68$ & $-0.03$ & $0.00$ & $-0.03$ & $-0.02$ & $-0.02$ & $0.00$ & $-0.01$ & $0.00$ & $0.00$ & $0.00$ & $0.00$ & $-0.31$ & 14$\to$14 & 13$\to$13 \\
 & $0.05$ & $+0.88$ & $-0.11$ & $-0.02$ & $-0.17$ & $-0.09$ & $-0.12$ & $0.00$ & $-0.03$ & $0.00$ & $-0.01$ & $0.00$ & $0.00$ & $+0.16$ & 14$\to$15 & 13$\to$15 \\
\bottomrule
\end{tabular}
\vspace{-2mm}
\end{table*}

Table~\ref{tab:main} reports how test-time optimization changes the $12$ metrics of the challenge for the seven systems that we analyze.
%Here, the reported values represent the difference between the modified signal $\xmod$ and the enhanced signal $\xhat$.
The shaded row of each system gives the scores of the enhanced signal $\xhat$, and the two rows below it give the difference of the modified signal $\xmod$ from these scores.
We jointly optimize the enhanced signal for the four non-intrusive MOS predictors, except in the rows marked (S), where we optimize the baseline's signal for DNSMOS alone.
We report two values of $\lambda$, i.e., $0.005$ and $0.05$.

The table shows that test-time optimization substantially increases the
predicted scores of the four MOS predictors simultaneously for all seven
systems at both values of $\lambda$. The other eight reference-based metrics generally remain unchanged
or decrease, except CAcc, which varies in both directions from $-1.34$ to $+0.43$ points. These results show that the improvement in the predicted scores
does not carry over to any of the other metrics. The weight $\lambda$ affects
the balance between the gain in the predicted scores and the degradation of
the other metrics. 
At $\lambda = 0.005$, the predicted scores rise less and the reference-based metrics move by no more than 0.01, whereas at $\lambda = 0.05$, the predicted scores rise further and the reference-based metrics degrade more.
%At $\lambda = 0.005$, the predicted scores rise less, and the other metrics move by no more than $0.01$, whereas at $\lambda = 0.05$ they rise further and the intrusive, and the downstream metrics degrade. 
This trend continues for larger values of $\lambda$, which we omit from the table for space.

The table also reports the predicted scores of Distill-MOS, an unseen predictor that is not included in the objective in Eq.~\eqref{eq:obj}. Although the four predicted scores rise substantially, the Distill-MOS score changes by at most $0.03$ in either direction at both values of
$\lambda$. 
These results show that the gain does not generalize to a predictor outside the objective, even though four different predictors are raised simultaneously.

In the rows marked (S), we can confirm that DNSMOS rises even more than under the joint optimization, whereas all other predicted scores decrease.
These results show that the four predictors disagree when only one of them is optimized, which is why we jointly optimize the enhanced signal for all of them.

\vspace{-1mm}
\subsection{Subjective evaluation}
\label{ssec:subj}

\begin{table}[t]
\centering
\setlength{\belowcaptionskip}{7pt}
\vspace{-2mm}
\caption{Differences in the MUSHRA scores, with a $95\%$ confidence interval
and the smallest equivalence bound at the $5\%$ level.}
\vspace{5pt}
\label{tab:subjective}
\setlength{\tabcolsep}{3pt}
\setlength{\aboverulesep}{0pt}
\setlength{\belowrulesep}{0pt}
\renewcommand{\arraystretch}{1.15}
\footnotesize
\begin{tabular}{l|cccc}
\toprule
Comparison\rule{0pt}{2.62ex}\rule[-1.08ex]{0pt}{0pt} & $\Delta$MUSHRA & 95\% CI & $p$ & Equiv. bound \\
\midrule
RICK$\,-\,$baird\rule{0pt}{2.62ex} & $+8.48$ & $[+2.5,+14.5]$ & $0.011$ & 13.4 \\
\midrule
baird, $\lambda\!=\!0.005$\rule{0pt}{2.62ex} & $-0.22$ & $[-2.2,+1.7]$ & $0.81$ & $1.8$ \\
baird, $\lambda\!=\!0.05$ & $-1.20$ & $[-5.0,+2.6]$ & $0.50$ & $4.3$ \\
RICK, $\lambda\!=\!0.005$\rule[-1.08ex]{0pt}{0pt} & $-0.65$ & $[-2.6,+1.3]$ & $0.47$ & $2.2$ \\
\bottomrule
\end{tabular}
\vspace{-2mm}
\end{table}

Table~\ref{tab:subjective} reports the results of the MUSHRA listening test.
The first row compares the original enhanced signals $\xhat$ of the two
systems, i.e., ``baird'' vs. ``RICK2000''. 
In the remaining rows, the modified signal $\xmod$ at the given $\lambda$ is compared with its original enhanced signal $\xhat$. 
Each $\Delta$ is the difference between the two conditions of the row, averaged over the $10$ utterances and the $11$ listeners. 
We report the $p$-value of a two-sided paired $t$-test~\cite{student1908probable} over the per-listener differences and the corresponding $95\%$ confidence interval (CI). 
In the test, the hidden reference obtained an average score of $99.9$ and the
anchor $1.6$, which confirms that the listeners performed the task as intended.
In the first row, $\Delta$MUSHRA is $+8.48$ points with $p = 0.011$, which
shows that the test works as intended, because it can detect a difference
between two different SE systems.

In the remaining rows, every $\Delta$MUSHRA is negative, i.e., none of the three optimized conditions is rated above its original signal.
Within ``baird'', it becomes more negative as $\lambda$ grows, although the predicted scores rise
further at the larger $\lambda$ (Table~\ref{tab:main}). These differences are
not significant, with $p = 0.81$, $0.50$, and $0.47$, but a non-significant
$p$-value does not support the absence of an effect. We thus apply two
one-sided tests (TOST) \cite{schuirmann1987comparison} and report the smallest equivalence bound at the $5\%$
level for each condition.
The bounds are $1.8$, $4.3$, and $2.2$ points on the $100$-point scale, e.g., the optimization of ``baird'' at $\lambda = 0.005$ changes perceived quality by at most $1.8$ points in either direction.
Even the largest of the three (i.e., $4.3$) is about half of the $8.48$-point difference between the two real systems in the first row.
These results suggest that test-time
optimization raises the predicted scores without improving perceived quality.

Incidentally, ``RICK2000'' was rated above ``baird'' in the first row, whereas the challenge ranked ``baird'' first.
Our test differs from the challenge's own listening test in the evaluated data and in the protocol, i.e., MUSHRA instead of MOS.
Therefore, we draw no conclusion about the relative
standing of the two systems. Our conclusions rest on the remaining rows, which compare each system with its modified version.

\vspace{-1mm}
\subsection{Impact on the challenge ranking}
\label{ssec:rank}

The last two columns of Table~\ref{tab:main} give the rank of each system on the leaderboard: the rank of $\xhat$ in the shaded rows and its change after the test-time optimization (before$\to$after) in the other rows. 
%before and after the test-time optimization.
Within the non-intrusive category, which contains the four MOS predictors in the objective, every system reaches rank $1$ or $2$ at $\lambda = 0.05$.
The overall rank rises as well, e.g., ``RICK2000'' from $6$ to $3$ at $\lambda = 0.005$ and the baseline from $13$ to $6$ at $\lambda = 0.05$. 
Under the rules of URGENT 2026, only the top six systems advance to the subjective evaluation.
Therefore, if test-time optimization were applied, even the baseline would qualify for the listening test on this leaderboard without any change to the SE model, displacing the system that held the sixth place.
We also observe that averaging the non-intrusive category with the three others reduces the effect on the overall rank, but does not remove it.

Baseline (S) shows the effect of the single-predictor optimization on the ranking.
Its rank does not improve at either value of $\lambda$, and it even falls both overall and within the non-intrusive category at $\lambda = 0.05$. These results show
that using multiple predictors in the ranking prevents the effect of optimizing for a single one~\cite{lin2022}, but not the effect of jointly optimizing for all of them.

Overall, the results observed in this paper suggest that test-time optimization, which directly increases the predicted scores, can distort the ranking of a challenge, moving a system up without any improvement in perceived quality. 
The extent of this effect would depend on what the objective ranking decides.
If it decides which systems are heard by listeners (e.g., URGENT 2026~\cite{urgent2026}), the effect is indirect. If it decides the final ranking (e.g., REAL-TSE 2026~\cite{realtse2026} and AVSE 2026~\cite{avse2026}), which can be a practical choice considering the time and financial costs of listening tests, the effect would be direct.
%In URGENT 2026, it decides which systems are heard by listeners, and the effect is therefore indirect.
%In challenges ranked by objective metrics that include MOS predictors, such as REAL-TSE 2026 [6] and AVSE 2026 [7], the effect would be direct.

%\vspace{-1mm}
\section{Conclusion}
\label{sec:conclusion}

In this paper, we analyzed test-time optimization with multiple MOS predictors, applied to systems submitted to the URGENT 2026 challenge. 
We found that the optimized predicted scores increased substantially, whereas none of the other measures improved correspondingly, including the reference-based metrics, downstream metrics, and an unseen predictor.
In a MUSHRA listening test, listeners rated the optimized signals no better than the original ones, which suggests that the predicted scores raised by test-time optimization do not correspond to better perceived quality. 
We also found that test-time optimization can move the challenge ranking by several places, which reveals a risk that it can distort the outcome of a challenge.
Using multiple predictors for ranking does not remove this risk: it can protect against optimizing only one of their predicted scores, but not against jointly optimizing all of their predicted scores.

We believe that these findings can inform evaluation practices in future SE research: they suggest that challenge organizers should keep the predictors used for ranking undisclosed and base their final decisions on listening tests.
In our future work, we plan to investigate how MOS predictors can be made robust against such optimization.

%\vspace{-1mm}
\section{Acknowledgements}

We thank the organizers of the URGENT 2026 challenge for providing the submitted signals, and all the participating teams for developing the SE systems; both enabled the analyses in this paper.

\vfill\pagebreak

\bibliographystyle{IEEEbib}
\bibliography{refs}

\begin{thebibliography}{10}

\bibitem{dnsmos}
Chandan K.~A. Reddy, Vishak Gopal, and Ross Cutler,
\newblock ``{DNSMOS} {P.835}: A non-intrusive perceptual objective speech
  quality metric to evaluate noise suppressors,''
\newblock in {\em IEEE International Conference on Acoustics, Speech and Signal
  Processing (ICASSP)}, 2022, pp. 886--890.

\bibitem{nisqa}
Gabriel Mittag, Babak Naderi, Assmaa Chehadi, and Sebastian M\"oller,
\newblock ``{NISQA}: A deep {CNN}-self-attention model for multidimensional
  speech quality prediction with crowdsourced datasets,''
\newblock in {\em Interspeech}, 2021, pp. 2127--2131.

\bibitem{utmos}
Takaaki Saeki, Detai Xin, Wataru Nakata, Tomoki Koriyama, Shinnosuke Takamichi,
  and Hiroshi Saruwatari,
\newblock ``{UTMOS}: {UTokyo-SaruLab} system for {VoiceMOS} challenge 2022,''
\newblock in {\em Interspeech}, 2022, pp. 4521--4525.

\bibitem{scoreq}
Alessandro Ragano, Jan Skoglund, and Andrew Hines,
\newblock ``{SCOREQ}: Speech quality assessment with contrastive regression,''
\newblock in {\em Advances in Neural Information Processing Systems (NeurIPS)},
  2024, vol.~37, pp. 105702--105729.

\bibitem{urgent2026}
Chenda Li, Wei Wang, Marvin Sach, Wangyou Zhang, Kohei Saijo, Samuele Cornell,
  et~al.,
\newblock ``{ICASSP} 2026 {URGENT} speech enhancement challenge,''
\newblock in {\em IEEE International Conference on Acoustics, Speech and Signal
  Processing (ICASSP)}, 2026, pp. 21919--21921.

\bibitem{realtse2026}
Shuai Wang, Zihan Qian, Ke~Zhang, Jiangyu Han, Zikai Liu, Xiaoyang Yu, et~al.,
\newblock ``{SLT} 2026 {REAL-TSE} challenge: Real-world target speaker
  extraction from conversational recordings,''
\newblock {\em arXiv preprint arXiv:2607.15198}, 2026.

\bibitem{avse2026}
Kai Li, Wenze Ren, Junjie Li, Cheng Yu, Peijun Yang, {Chien-yu} Huang, et~al.,
\newblock ``The {ISCSLP} 2026 real-world audio-visual speech enhancement
  challenge,''
\newblock {\em arXiv preprint arXiv:2608.23759}, 2026.

\bibitem{multimetric2025}
Wei Wang, Wangyou Zhang, Chenda Li, Jiatong Shi, Shinji Watanabe, and Yanmin
  Qian,
\newblock ``Improving speech enhancement with multi-metric supervision from
  learned quality assessment,''
\newblock in {\em IEEE Automatic Speech Recognition and Understanding Workshop
  (ASRU)}, 2025, pp. 1--8.

\bibitem{matsunaga2026three}
Ryutaro Matsunaga, Ryo Takahashi, and Shinnosuke Takamichi,
\newblock ``Three-stage {BSRNN} for universal speech enhancement and data
  curation using a large pre-trained speech restoration model,''
\newblock in {\em IEEE International Conference on Acoustics, Speech and Signal
  Processing (ICASSP)}, 2026, pp. 21904--21906.

\bibitem{fgsm}
Ian~J. Goodfellow, Jonathon Shlens, and Christian Szegedy,
\newblock ``Explaining and harnessing adversarial examples,''
\newblock in {\em International Conference on Learning Representations (ICLR)},
  2015.

\bibitem{lin2022}
Hsin-Yi Lin, Huan-Hsin Tseng, and Yu~Tsao,
\newblock ``On the robustness of non-intrusive speech quality model by
  adversarial examples,''
\newblock in {\em IEEE International Conference on Acoustics, Speech and Signal
  Processing (ICASSP)}, 2023, pp. 1--5.

\bibitem{huang2026}
Wen-Chin Huang and Tomoki Toda,
\newblock ``Attacking {UTMOS}: Probing the robustness of a speech quality
  assessment model,''
\newblock {\em arXiv preprint arXiv:2606.31105}, 2026.

\bibitem{merl2026}
Dominik Klement, Yoshiki Masuyama, Christoph Boeddeker, Kohei Saijo, Julius
  Richter, Gordon Wichern, et~al.,
\newblock ``Technical report for {MERL}'s {Real-TSE} challenge submission,''
  Real-TSE Challenge official system description,
  \url{https://real-tse.github.io/assets/pdf/MERL-SA-Track2.pdf}, 2026.

\bibitem{rong2026gap}
Xiaobin Rong, Yushi Wang, Zheng Wang, and Jing Lu,
\newblock ``{GAP-URGENet}: A generative-predictive fusion framework for
  universal speech enhancement,''
\newblock in {\em IEEE International Conference on Acoustics, Speech and Signal
  Processing (ICASSP)}, 2026, pp. 21895--21897.

\bibitem{goswami2026hybrid}
Nabarun Goswami and Tatsuya Harada,
\newblock ``Hybrid speech enhancement with discriminative and codec token
  prediction models guided by cleaned {SSL} features for the {ICASSP} 2026
  {URGENT} challenge,''
\newblock in {\em IEEE International Conference on Acoustics, Speech and Signal
  Processing (ICASSP)}, 2026, pp. 21916--21918.

\bibitem{liu2026hybrid}
Yinghao Liu, Chengwei Liu, Xiaotao Liang, Haoyin Yan, Shaofei Xue, and Zheng
  Xue,
\newblock ``A hybrid discriminative and generative system for universal speech
  enhancement,''
\newblock in {\em IEEE International Conference on Acoustics, Speech and Signal
  Processing (ICASSP)}, 2026, pp. 21907--21909.

\bibitem{urgentcode}
``{URGENT} 2026 challenge {track}~1,''
  \url{https://github.com/urgent-challenge/urgent2026_challenge_track1}, 2026.

\bibitem{pesq}
Antony~W. Rix, John~G. Beerends, Michael~P. Hollier, and Andries~P. Hekstra,
\newblock ``Perceptual evaluation of speech quality {(PESQ)}---a new method for
  speech quality assessment of telephone networks and codecs,''
\newblock in {\em IEEE International Conference on Acoustics, Speech and Signal
  Processing (ICASSP)}, 2001, vol.~2, pp. 749--752.

\bibitem{estoi}
Jesper Jensen and Cees~H. Taal,
\newblock ``An algorithm for predicting the intelligibility of speech masked by
  modulated noise maskers,''
\newblock {\em IEEE/ACM Transactions on Audio, Speech, and Language Processing
  (TASLP)}, vol. 24, no. 11, pp. 2009--2022, 2016.

\bibitem{sbertscore}
Takaaki Saeki, Soumi Maiti, Shinnosuke Takamichi, Shinji Watanabe, and Hiroshi
  Saruwatari,
\newblock ``{SpeechBERTScore}: Reference-aware automatic evaluation of speech
  generation leveraging {NLP} evaluation metrics,''
\newblock in {\em Interspeech}, 2024, pp. 4943--4947.

\bibitem{lps}
Jan Pirklbauer, Marvin Sach, Kristoff Fluyt, Wouter Tirry, Wafaa Wardah,
  Sebastian M\"{o}ller, et~al.,
\newblock ``Evaluation metrics for generative speech enhancement methods:
  Issues and perspectives,''
\newblock in {\em Speech Communication; 15th ITG Conference}, 2023, pp.
  265--269.

\bibitem{distillmos}
Benjamin Stahl and Hannes Gamper,
\newblock ``Distillation and pruning for scalable self-supervised
  representation-based speech quality assessment,''
\newblock in {\em IEEE International Conference on Acoustics, Speech and Signal
  Processing (ICASSP)}, 2025, pp. 1--5.

\bibitem{bs1534}
{ITU-R},
\newblock ``Recommendation {ITU-R} {BS}.1534-3: Method for the subjective
  assessment of intermediate quality level of audio systems,'' International
  Telecommunication Union, 2015.

\bibitem{multiresl1}
Yen-Ju Lu, Samuele Cornell, Xuankai Chang, Wangyou Zhang, Chenda Li, Zhaoheng
  Ni, et~al.,
\newblock ``Towards low-distortion multi-channel speech enhancement: The
  {ESPNet-SE} submission to the {L3DAS22} challenge,''
\newblock in {\em IEEE International Conference on Acoustics, Speech and Signal
  Processing (ICASSP)}, 2022, pp. 9201--9205.

\bibitem{student1908probable}
Student,
\newblock ``The probable error of a mean,''
\newblock {\em Biometrika}, vol. 6, no. 1, pp. 1--25, 1908.

\bibitem{schuirmann1987comparison}
Donald~J. Schuirmann,
\newblock ``A comparison of the two one-sided tests procedure and the power
  approach for assessing the equivalence of average bioavailability,''
\newblock {\em Journal of Pharmacokinetics and Biopharmaceutics}, vol. 15, no.
  6, pp. 657--680, 1987.

\end{thebibliography}

\end{document}